\documentclass[sigconf]{acmart}
\usepackage{adjustbox}
\usepackage{paralist}
\usepackage{multirow,multicol}
\usepackage{microtype}
\usepackage{booktabs}
\usepackage{color}
\usepackage{listings}
\definecolor{green2}{RGB}{123, 163, 101}
\definecolor{yellow2}{RGB}{196, 174, 137}
\definecolor{cyan2}{RGB}{123, 163, 101}
\definecolor{pink2}{RGB}{123, 163, 101}
\definecolor{brown2}{RGB}{204, 163, 122}
\AtBeginDocument{%
  \providecommand\BibTeX{{%
    \normalfont B\kern-0.5em{\scshape i\kern-0.25em b}\kern-0.8em\TeX}}}

\setcopyright{acmlicensed}
\copyrightyear{2018}
\acmYear{2018}
\acmDOI{XXXXXXX.XXXXXXX}

\acmConference[Conference acronym 'XX]{Make sure to enter the correct
  conference title from your rights confirmation emai}{June 03--05,
  2018}{Woodstock, NY}
\acmBooktitle{Woodstock '18: ACM Symposium on Neural Gaze Detection,
 June 03--05, 2018, Woodstock, NY} 
\acmISBN{978-1-4503-XXXX-X/18/06}

\begin{document}

\title{Efficiently and Effectively: A Two-stage Approach to Balance Plaintext and Encrypted Text for Traffic Classification}

\author{Wei Peng}
\orcid{0000-0001-8179-1577}
\affiliation{%
	\institution{Zhongguancun Laboratory}
	\city{Beijing}
	\country{China}
}

\author{Lei Cui}
\affiliation{%
	\institution{Zhongguancun Laboratory}
	\city{Beijing}
	\country{China}
}

\author{Wei Cai}
\affiliation{%
	\institution{Zhongguancun Laboratory}
	\city{Beijing}
	\country{China}}

\author{Zhenquan Ding}
\affiliation{%
	\institution{Zhongguancun Laboratory}
	\city{Beijing}
	\country{China}}

\author{Zhiyu Hao*}
\affiliation{%
	\institution{Zhongguancun Laboratory}
	\city{Beijing}
	\country{China}}

\author{Xiaochun Yun}
\authornote{Corresponding author.}
\affiliation{%
	\institution{Zhongguancun Laboratory}
	\city{Beijing}
	\country{China}}

\renewcommand{\shortauthors}{Trovato and Tobin, et al.}

\begin{abstract}
  Encrypted traffic classification is the task of identifying the application or service associated with encrypted network traffic. One effective approach for this task is to use deep learning methods to encode the raw traffic bytes directly and automatically extract features for classification (byte-based models). However, current byte-based models input raw traffic bytes, whether plaintext or encrypted text, for automated feature extraction, neglecting the distinct impacts of plaintext and encrypted text on downstream tasks. Additionally, these models primarily focus on improving classification accuracy, with little emphasis on the efficiency of models. In this paper, for the first time, we analyze the impact of plaintext and encrypted text on the model's effectiveness and efficiency. Based on our observations and findings, we propose an efficient and effective two-stage approach to balance the trade-off between plaintext and encrypted text in traffic classification. Specifically, stage one proposes a DPC selector to Determine whether the Plaintext information is sufficient to perform subsequent Classification (DPC). This stage quickly identifies samples that can be classified using plaintext, leveraging explicit byte features in plaintext to enhance model's efficiency. Stage two aims to adaptively make a classification with the result from stage one. This stage could incorporate encrypted text information for samples that cannot be classified using plaintext alone, ensuring the model's effectiveness on traffic classification tasks. Experiments on two public datasets and one real-world collected dataset demonstrate that our proposed model achieves state-of-the-art results in both effectiveness and efficiency. The model code will be released in https://github/after/review.
\end{abstract}





\maketitle

\section{Introduction}

\begin{figure}[t]
	\centering
	\includegraphics[width=0.47\textwidth]{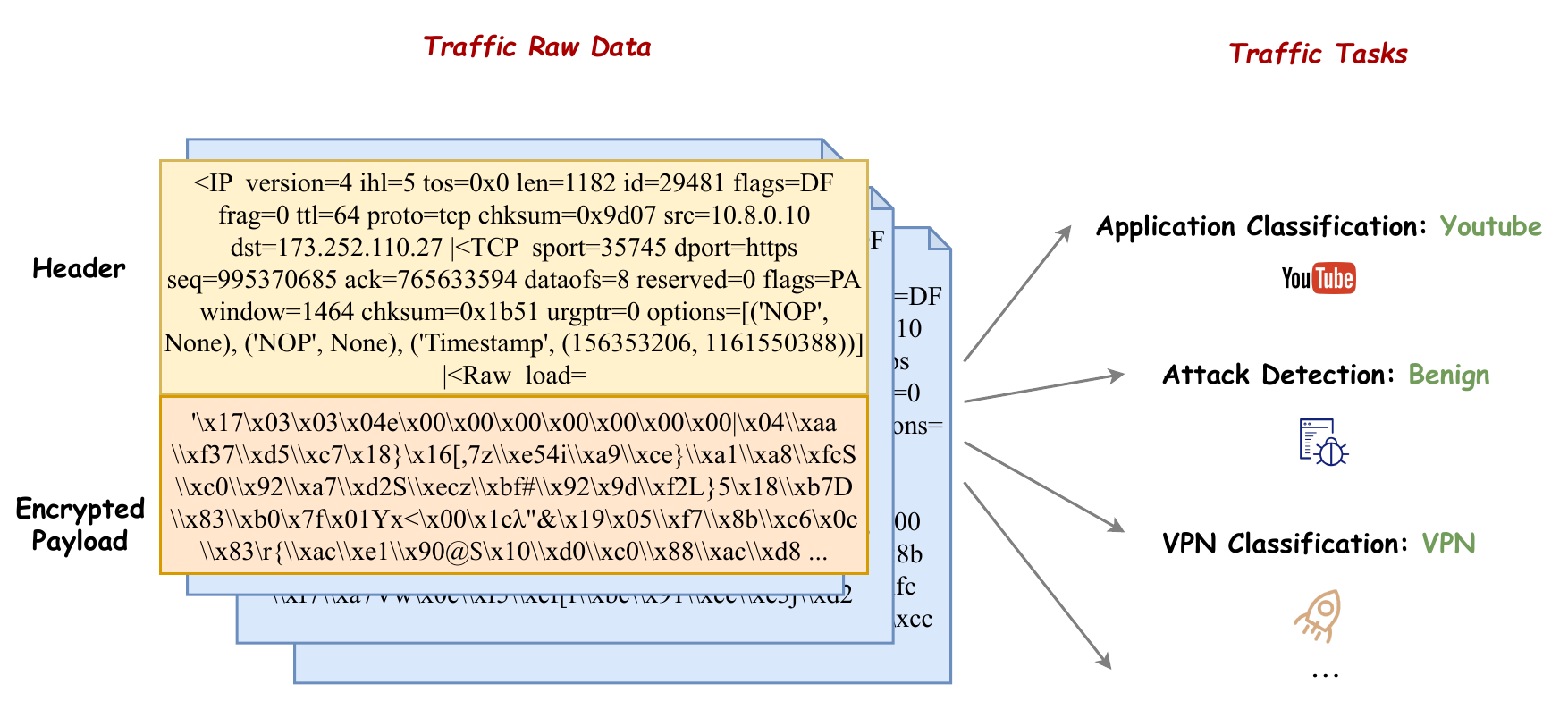}
	\caption{An example of traffic data packet. The right-hand part means the downstream traffic classification task. {\color{green2} Green} font represents the different labels.
	}
	\label{fig:example}
\end{figure}

Encrypted traffic classification is a critical task in the field of cybersecurity, focusing on identifying and categorizing network traffic data that has been encrypted \cite{DBLP:journals/cn/BujlowCB15,DBLP:journals/ijon/AcetoCMP20}. This process is essential for maintaining network security, managing bandwidth, detecting malicious activities, and ensuring regulatory compliance \cite{DBLP:journals/pomacs/AkbariSVLBMMT21,DBLP:conf/issta/0003CJHLD23}. Applications of encrypted traffic classification span various domains, including intrusion detection systems, quality of service (QoS) management, and network forensics. As illustrated in Figure \ref{fig:example}, some common applications include encrypted application classification \cite{DBLP:conf/www/LinXGLSY22}, attack detection \cite{DBLP:journals/ton/MengWWYZ21}, VPN classification \cite{DBLP:conf/icissp/Draper-GilLMG16,DBLP:conf/www/OhL0BK23}, etc.

Network traffic is composed of packets, as shown in Figure \ref{fig:example}, each consisting of two main components: the header (plaintext) and the payload (usually encrypted text). The header contains metadata about the packet, such as source and destination IP addresses, protocol information, and so on. The payload, on the other hand, contains the actual data being transmitted and is often encrypted in the form of hexadecimal to ensure confidentiality and integrity. Current byte-based encrypted traffic classification researches \cite{DBLP:conf/infocom/LiuHXCL19,DBLP:journals/comcom/JemalHCM21,DBLP:conf/kdd/MengWMLLZ22,zhao2023yet} typically use both the header and the payload as inputs to the models. A fixed number of bytes, like 128, 256, 512, etc., is utilized for the classification task. These bytes often include a mix of plaintext information from the header and encrypted data from the payload. At the same time, some studies \cite{DBLP:journals/cacm/AkbariSVLBMMT22,DBLP:conf/www/ZhangYXLMLL23} have shown that plaintext information significantly contributes to the effectiveness of classification. For example, the paper \cite{DBLP:journals/cacm/AkbariSVLBMMT22} mentions that \textbf{the availability of some obvious feature of the server (e.g., plain-text SNI field) in-the-clear is crucial for the utility of deep-learning-based (DL) approaches}. Another research \cite{DBLP:conf/www/ZhangYXLMLL23} proves through ablation experiments that \textbf{plaintext information is more helpful for the effectiveness of classification than encrypted text information}. This leads to a fascinating hypothesis: Can we achieve satisfactory traffic classification performance using only the plaintext information present in the header? 

\begin{figure*}[t]
	\centering
	\includegraphics[width=0.98\textwidth]{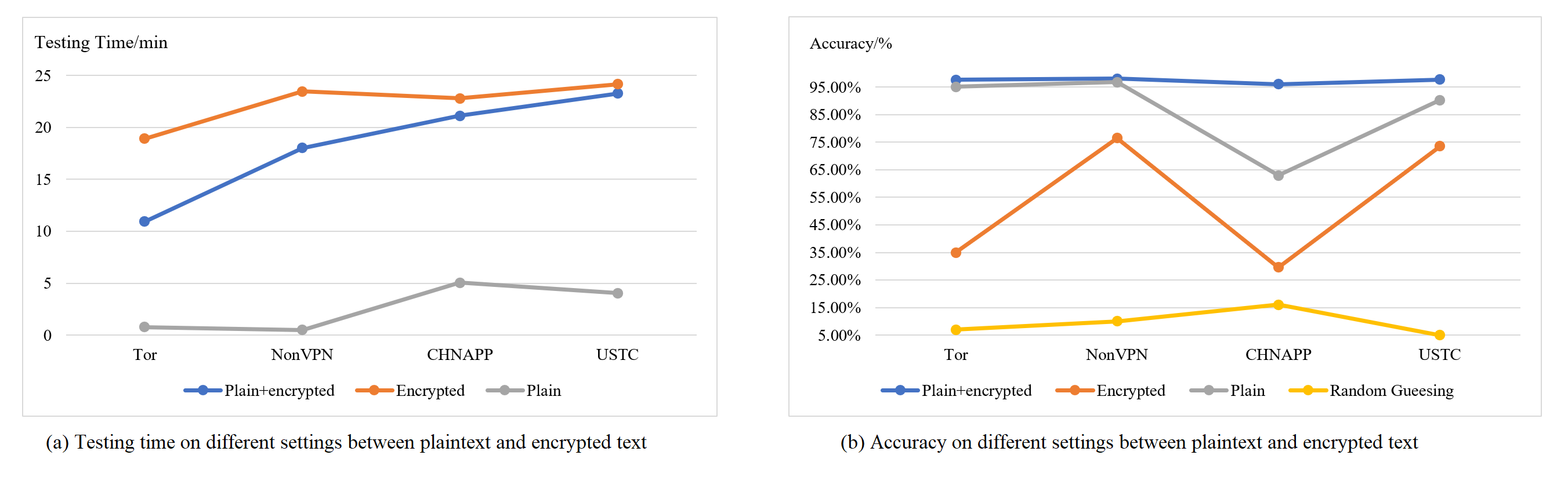}
	\caption{Testing time and accuracy analysis on different settings between plaintext and encrypted text on four datasets.
	}
	\label{fig:example-s1}
\end{figure*}
To further validate the roles of plaintext, encrypted text, and their combination in traffic classification, we compare the one of the state-of-the-art (SOTA) models \cite{DBLP:conf/kdd/MengWMLLZ22} performance and time overhead across four datasets \cite{DBLP:conf/icissp/Draper-GilLMG16,DBLP:conf/icissp/LashkariDMG17,DBLP:conf/icoin/WangZZYS17}. For the combined plaintext-encrypted text setting, as shown in Figure 2(b), the current prevalent methods that utilize both plaintext and encrypted text achieve the highest accuracy (indicated by the blue line). However, this setting results in high time overhead, as depicted in Figure 2(a) blue line.

Some studies \cite{DBLP:journals/cacm/AkbariSVLBMMT22,DBLP:conf/www/ZhangYXLMLL23} have demonstrate that the effectiveness of DL-based methods largely rely on plaintext information. This suggests that satisfactory classification performance can be achieved by leveraging only plaintext data. Notably, for the plaintext-only setting, as indicated by the gray line in Figure 2(b), the model achieves around 90\% accuracy on certain datasets, further substantiating the crucial role of plaintext in classification. Surprisingly, this setting also enhances time efficiency by approximately 4 to 10 times (indicated by the gray line in Figure 2(a)). However, relying solely on plaintext data obtains lower accuracy on some situations, like result on USTC and CHNAPP datasets \cite{DBLP:conf/icoin/WangZZYS17} in Figure 2(b).

For the encrypted text setting, the results in Figure 2(b) indicate that using purely encrypted text for classification performs better than random guessing, implying that encrypted data can also aid in improving classification accuracy alongside plaintext.

In conclusion, the time overhead for the combined plaintext-encrypted text setting is considerably high, although the accuracy is the best in these settings. Therefore, balancing the utilization of plaintext and encrypted text to ensure accurate classification while optimizing time efficiency remains a key research challenge. \textbf{Our study aims to explore this trade-off to achieve both high accuracy and low time overhead in traffic classification.}

To address the above issues, we propose a simple but Efficient and Effective Two-Stage approach (EETS) to balance plaintext and encrypted text for traffic classification. In the first stage, we introduce a new traffic classification task: Determining whether the Plaintext is sufficient to perform subsequent Classification (DPC) and propose a DPC selector. The purpose of this stage is to fully leverage plaintext information to quickly determine if the traffic can be classified using only plaintext (plaintext classifiable). This allows us to enhance the time efficiency by avoiding unnecessary processing of encrypted data when plaintext alone suffices. In the second stage, we proceed with the plaintext classification model for a efficient classification if the traffic can be classified only using plaintext. Otherwise (plaintext non-classifiable), a combination of plaintext and encrypted text is considered to further improve the model's accuracy.
The two-stage design ensures a balance between effectiveness and efficiency. By attempting to classify with plaintext, DL-based models can significantly reduce processing time for a large portion of the traffic. For cases where plaintext is insufficient, incorporating encrypted text ensures that models still achieve high accuracy, while maintaining high time efficiency.

The contributions can be summarized as follows:
\begin{compactitem}
	\item For the first time, we analyze the impact of plaintext and encrypted text on model performance and time overhead. Based on this, we introduce an interesting idea of categorizing data packets into two types: plaintext classifiable and plaintext non-classifiable. Then, a simple but efficient and effective approach is designed to significantly reduce time overhead while maintaining high classification accuracy.
	\item We propose a two-stage approach to balance the trade-off between plaintext and encrypted text in traffic classification. Stage one proposes a DPC selector to determine whether the plaintext information is sufficient to perform subsequent classification. Stage two is to adaptively make a classification with the result from stage one.
	\item Experimental results on two public datasets and one real-world dataset collected by ourselves demonstrate the efficiency and effectiveness of our approach.
\end{compactitem}

\begin{figure*}[!]
	\centering
	\includegraphics[width=0.98\textwidth]{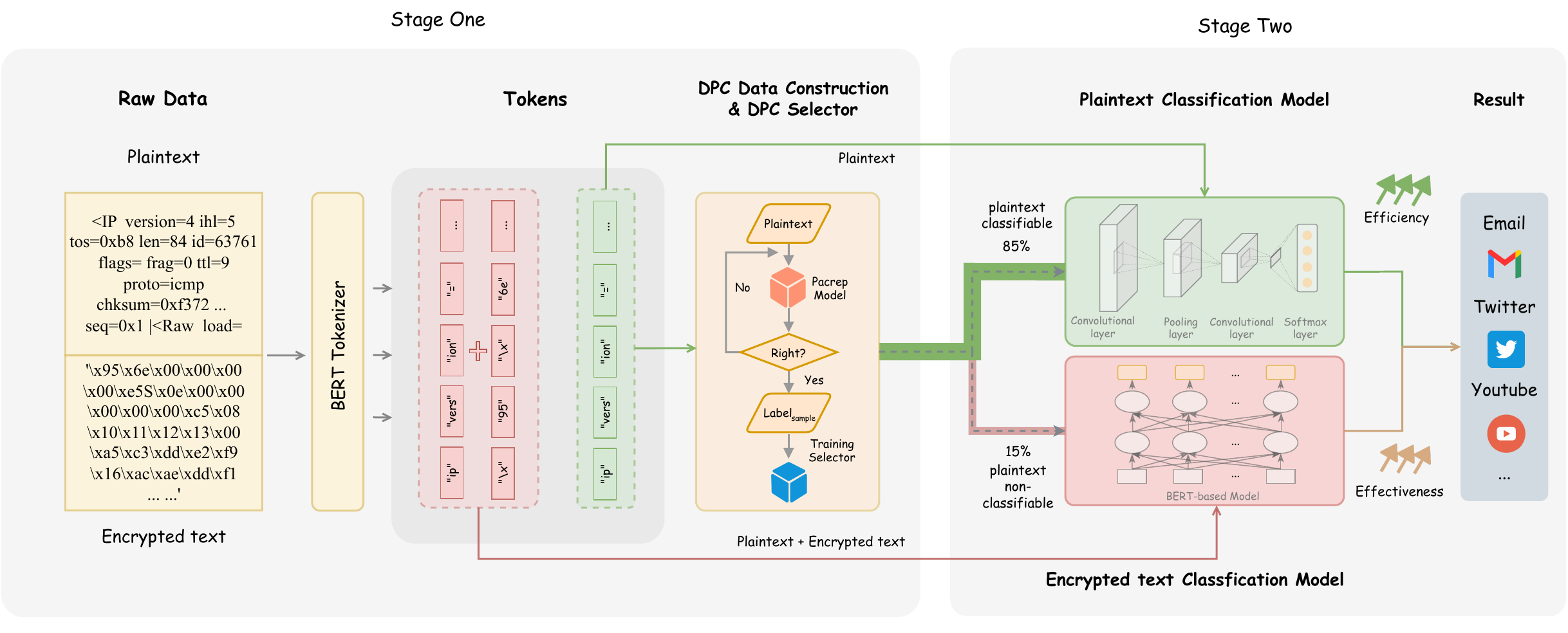}
	\caption{The overview of EETS, which consists of two stages. Stage one proposes a DPC selector to determine whether the plaintext is sufficient to perform subsequent classification. Stage two is to adaptively make a classification with the result from stage one. The {\color{green2} green} arrows indicate the plaintext processing, {\color{red} red} arrows present the plaintext and encrypted text processing.}
	\label{fig:model}
\end{figure*}

\section{Related Work}

\subsection{Plaintext Classification}
In the early days of network traffic classification, methods \cite{cheng2011traffic,DBLP:conf/apnoms/YoonPPOK09,DBLP:conf/apnoms/ZhangMWL14,DBLP:conf/im/ShbairCGC15,DBLP:conf/icdcsw/ShbairCFC16} mainly rely on plaintext information to analyze traffic data. This includes using transport layer protocol port numbers, such as UDP or TCP, or leveraging plaintext information like the Server Name Indication (SNI). Port-based methods \cite{cheng2011traffic,DBLP:conf/apnoms/YoonPPOK09,DBLP:conf/apnoms/ZhangMWL14} are straightforward to implement and have low time complexity, making them a popular choice when only specific port applications needed to be classified. However, as applications and protocols diversified and techniques such as port hopping and port masquerading emerge, the accuracy of port-based classification methods decrease, rendering them unreliable. SNI-based methods \cite{DBLP:conf/im/ShbairCGC15,DBLP:conf/icdcsw/ShbairCFC16} use the hostname information requested during the TLS handshake to determine the traffic category. These methods boast high accuracy and are easy to implement. Nevertheless, not all clients support SNI, particularly older browsers or operating systems. When such clients attempt to access SNI-enabled websites, SNI-based methods fail. 

Rencently, with the continuous evolution of encryption protocols \cite{DBLP:journals/ton/HanKCCHH20,DBLP:conf/aiiot/ThakurHTKA23}, plaintext classification models have become increasingly limited. Early plaintext classification approaches heavily relied on manually crafted feature engineering, and solely using plaintext information may not suffice for certain scenarios. As a result, researchers have shifted towards DL-based models, which allow for automatic feature extraction without the need to restrict the data to plaintext or encrypted text. This approach enhances the adaptability and robustness of network traffic classification methods.

\subsection{Encrypted Traffic Classification}
From a methodological perspective, encrypted traffic classification techniques can be broadly categorized into feature-based methods and byte-based models. Given that this paper directly uses both the header and payload content as inputs, the focus will be on byte-based models \cite{DBLP:conf/ccs/SirinamIJW18,DBLP:conf/infocom/LiuHXCL19,DBLP:journals/comcom/JemalHCM21,DBLP:conf/kdd/MengWMLLZ22,zhao2023yet}. The details are as follows.

Byte-based models leverage raw bytes sequences of network traffic for classification. These models do not rely on manually crafted features but utilize the inherent information present in the raw bytes. This approach allows the models to automatically learn the relevant patterns and features necessary for accurate classification, making them highly suitable for handling encrypted traffic where traditional feature extraction methods fall short.

Byte-based models \cite{DBLP:conf/ccs/SirinamIJW18,DBLP:conf/infocom/LiuHXCL19,DBLP:journals/comcom/JemalHCM21,DBLP:conf/kdd/MengWMLLZ22,zhao2023yet} work directly on raw traffic bytes and learn representations using techniques like Convolutional Neural Networks (CNNs) and Long Short-Term Memory Networks (LSTMs). They can capture complex patterns in encrypted data packets without manual feature engineering. These approaches can be categorized as image-based or text-based depending on how they interpret the input traffic. Specifically, image-based models \cite{DBLP:conf/ccs/SirinamIJW18,DBLP:journals/comcom/JemalHCM21,zhao2023yet} treat the traffic data as 2D images by transforming the raw bytes into grayscale intensity values or RGB colors. They then apply well-known image classification CNN architectures like ResNet and Inception that have shown excellent performance in computer vision. These models can identify visual patterns in the transformed input matrix based on learned convolutional filters. Text-based models such as Recurrent Neural Networks (RNN) and LSTMs process the input sequentially as text sequences \cite{DBLP:journals/soco/LotfollahiSZS20,DBLP:conf/kdd/MengWMLLZ22}. They can analyze long-range dependencies in the raw traffic bytes. Self-attention models like transformer are also popular for learning contextual relationships in encrypted traffic data. For example, \citet{DBLP:conf/kdd/MengWMLLZ22} utilize contrastive loss with a sample selector and attention mechanism to optimize the learned representations with various traffic classification tasks. 

\section{Models}
The proposed approach EETS is presented in Figure \ref{fig:model}, which efficiently and effectively to balance plaintext and encrypted text for traffic classification. The core modules consist of a DPC selector (in stage one) and plaintext or encrypted text classification models (in stage two). Specifically, the DPC selector is able to determine whether the plaintext information is sufficient to perform subsequent classification. The purpose of this module is to fully leverage plaintext information to quickly determine if the traffic can be classified using only plaintext. Then, plaintext or encrypted text classification models can make an adaptively classification with the result from stage one. In the following, we first describe the problem formulation. Then, the detail of the EETS is introduced.

\subsection{Problem Formulation}
Given a pcap file, the scapy toolkit parses the file and extracts the network packets. Each packet includes header portion (plaintext) $X^p$ and payload data (usually encrypted text) $X^e$ in the form of string $X = \{X^p, X^e\} = \{x^p_{1}, x^p_{2}, \dots, x^p_{M}, \dots, x^e_{N}\}$ that consists of ${N}$ traffic words, with a category label $y$, e.g. YouTube, Facebook, etc., where $p$ indicates plaintext, $e$ indicates encrypted text, $M$ means the number of plaintext traffic words. Our task can be divided into two subtasks. The first subtask involves determining, based on plaintext information, whether the data is sufficient for subsequent classification. 
The second subtask depends on the result of the first subtask, selecting different branches to output the corresponding packet category label, which is a multi-class classification task.

\subsection{Training Stage One}
\subsubsection{DPC Data Construction} 
In the first stage of training, the proposed new DPC task involves assessing whether the plaintext information is sufficient for subsequent classification. However, existing datasets do not contain labels for this specific purpose, making the construction of the DPC data a crucial step.

Considering that the DPC task is a traffic classification problem and relatively straightforward, many SOTA traffic classification models \cite{DBLP:conf/www/LinXGLSY22,DBLP:conf/kdd/MengWMLLZ22} can be leveraged for data labeling. Specifically, we use only plaintext information as input and employ the Pacrep model to perform standard traffic classification training. Once the model is trained, we predict the labels for all data samples. A data sample is labeled as ``1'' (plaintext classifiable) only if the predicted label matches the original sample's label. Otherwise, the sample is labeled as ``0'' (plaintext non-classifiable). The formulation can be defined as follows:
\begin{equation} \label{equ:1}
	label_{{\rm{sapmle}}} = \begin{cases}
		1,  & \text{if $y$ == $\hat{y}$} \\
		0, & \text{if $y$ ~!= $\hat{y}$}
	\end{cases}
\end{equation}
where $\hat{y}$ is the predicted label by the traffic classification model.

After obtaining all the labels, we can then train the DPC selector, which will enable us to distinguish between plaintext classifiable and non-classifiable effectively. This process ensures accurate and efficient traffic classification in subsequent stage.

\subsubsection{DPC Selector} 
After obtaining the training data, we proceed to train the DPC selector. This module aims to filter traffic data, directing plaintext classifiable samples to the plaintext classification model and plaintext non-classifiable samples to the encrypted classification model. This adaptive approach enhances the effectiveness of classification for different types of data, as well as playing a crucial role in optimizing the efficiency of traffic classification. Specifically, existing off-the-shelf traffic classification model \footnote{https://github.com/ict-net/PacRep} is utilized as the DPC selector to ensure robust and effective performance. The cross-entropy loss function is used for training. Then, we test the model to selectively separate the samples in the test set into two categories. As illustrated in Figure \ref{fig:model}, our experiments show that the majority of data (represented by the green portion) is directed towards the plaintext classification model, while a smaller portion is directed towards the encrypted text classification model. This demonstrates the superiority of our proposed framework, effectively and efficiently achieving traffic classification.

\begin{figure}[t]
	\centering
	\includegraphics[width=0.35\textwidth]{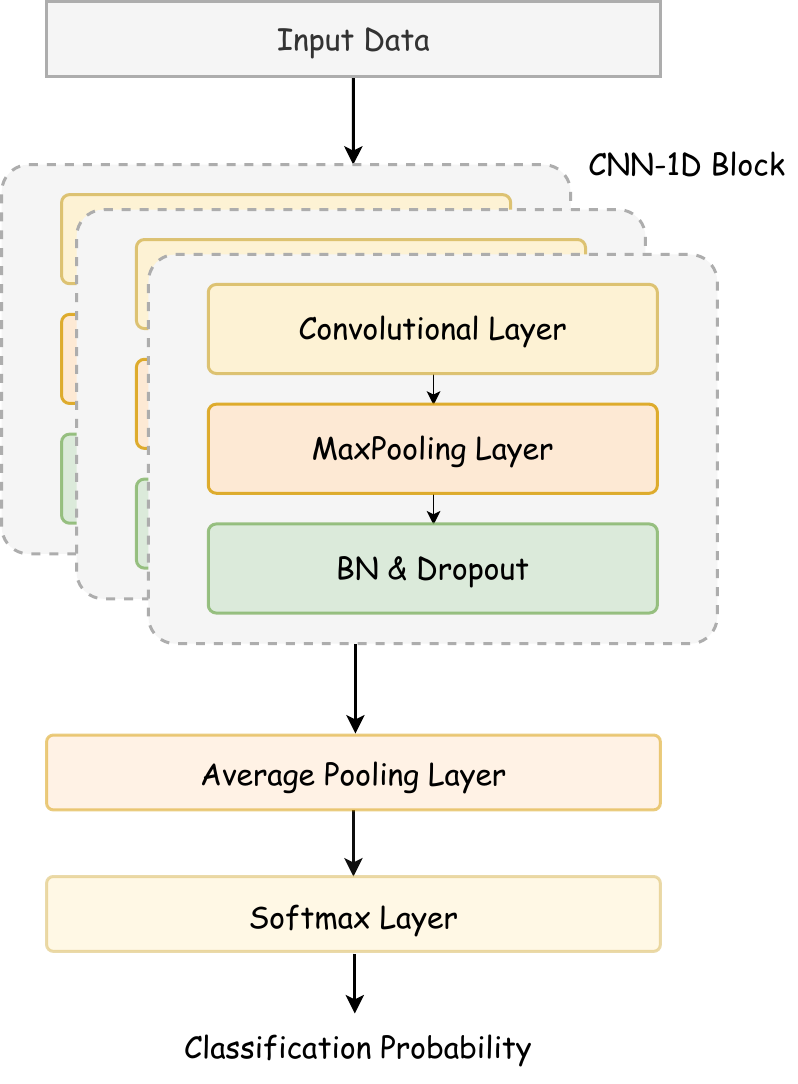}
	\caption{An overview of plaintext classification model.
	}
	\label{fig:model-1}
\end{figure}

\subsection{Training Stage Two}
In the second phase, the proposed EETS aims to adaptively select different branches for traffic classification, ensuring both effectiveness and efficiency. To further enhance the model's efficiency, the plaintext classification model will employ a simple CNN-based deep learning approach, which allows for faster predictions. On the other hand, the encrypted classification model will utilize more complicated DL-based techniques, like attention mechanism or Pre-trained Language Model (PLM), to ensure high accuracy. Below, we provide a detailed introduction to these two modules.

\subsubsection{Plaintext Classification Model} 
The plaintext classification model leverages CNNs to enhance the model's efficiency, which are well-suited to quickly process and classify a large number of data. CNNs are advantageous in this context due to their efficiency in feature extraction and their ability to capture contextual relationships in the data. This allows the model to make fast and accurate predictions, making it ideal for scenarios where rapid decision-making is critical. Specifically, the model begins by tokenizing the input plaintext to obtain a sequence of strings. These words are then converted into corresponding numerical values using the one-hot encoding layer, resulting in a sequence $\{h_{1}, h_{2}, \dots, h_{M}\}$:
\begin{equation} \label{equ:2}
	\{h_{1}, h_{2}, \dots, h_{M}\} = {\rm{One\verb|-|hot ~ Layer}} ~ (\{x^p_{1}, x^p_{2}, \dots, x^p_{M}\})
\end{equation}
where $h_m \in \mathbb{R}^{1 \times k}$, $k$ is the dimension of one-hot layer, $m$ is the index, $M$ is the total number of plaintext words.

This numerical representation enables the model to perform subsequent computations. Next, a 1D-convolution layer with a dimension of (1, $d_c$) is employed to capture local dependencies within the sequence and extract essential features. The advantage of using a 1D-convolution layer lies in its ability to efficiently process sequential data by sliding a filter across the input, thus detecting patterns and relationships within a localized context. The encoded hidden vectors are as follows:
\begin{equation} \label{equ:3}
	\{\tilde{h_{1}}, \tilde{h_{2}}, \dots, \tilde{h_{M}}\} = {\rm{1D\verb|-|convolution ~ Layer}} ~ (\{h_{1}, h_{2}, \dots, h_{M}\})
\end{equation}
where $\tilde{h_m} \in \mathbb{R}^{1 \times d_c}$, $d_c$ is the dimension of 1D-convolution layer.

Following this, max-pooling is applied to reduce the dimensionality of the hidden vectors and highlight the most significant features, enhancing computational efficiency and robustness to variations in the input. Batch normalization and dropout are then utilized to normalize the activations and prevent overfitting, respectively. Finally, average pooling is performed before passing the features through a softmax layer to output the probability distribution $p \in \mathbb{R}^{l}$ over the classes, where $l$ is the total number of applications in the dataset. This distribution indicates the likelihood of the input belonging to each classes.

\subsubsection{Encrypted Text Classification Model} 
For the encrypted classification model, a more intricate DL-based method is employed to handle the additional complexity of encrypted data. Techniques such as attention mechanism or Transformer-based models, e.g.BERT \cite{DBLP:conf/naacl/DevlinCLT19}, can be considered for their superior ability to manage sequential and highly structured data. These models, though computationally intensive, provide the necessary accuracy to classify encrypted traffic effectively. At the same time, after the first selecting stage, only a smaller portion of the data is towards this branch. Specifically, given the complexity of encrypted text, we utilize pre-trained language models such as BERT \cite{DBLP:conf/naacl/DevlinCLT19,DBLP:conf/www/LinXGLSY22} as the encoder in our encrypted classification model. The BERT is employed to encode network traffic data, resulting in rich hidden representations. The choice of BERT leverages its capability to understand and process complex traffic patterns, which is crucial for effectively handling encrypted traffic data, the equation is defined as:
\begin{equation} \label{equ:bert}
	\{h_{1}, h_{2}, \dots, h_{N}\} = {\rm{BERT}} ~ (\{x^p_{1}, x^p_{2}, \dots, x^e_{N}\})
\end{equation}
where $h_n \in \mathbb{R}^{1 \times d_b}$, $d_b$ is the dimension of BERT layer, $n$ is the index, $N$ is the total number of traffic words.

Subsequently, we apply an attention mechanism to model the interactions between different parts of the encoded traffic data. The attention mechanism allows the model to focus on the most relevant portions of the data, enhancing the model's ability to capture intricate dependencies within the traffic sequences. The updated hidden representation can be expressed in the following:
\begin{equation} \label{equ:att}
	\{\tilde{h_{1}}, \tilde{h_{2}}, \dots, \tilde{h_{N}}\} = {\rm{Softmax}} ~ (\frac{QK^T}{\sqrt{d_b}})V
\end{equation}
where $Q = K = V = \{h_{1}, h_{2}, \dots, h_{N}\}$, and $\tilde{h_n}$ denotes the refined hidden representations post-attention.

Finally, these refined hidden representations are passed through a classifier, e.g. multilayer perceptron, which outputs the probability distribution over the possible traffic classes, as:
\begin{equation} \label{equ:att3}
	P(y|\tilde{H}) = {\rm{Softmax}}~({\rm{Classifier}}~(\{\tilde{h_{1}}, \tilde{h_{2}}, \dots, \tilde{h_{N}}\}))
\end{equation}

\begin{table}[t]
	\centering
	\caption{Statistics of the datasets, including two public datasets and one collected real-world dataset.}
	\label{tab:stats}
	\begin{tabular}{lcccc}
		\toprule
		\textbf{Dataset}                         & \textbf{Total Packets}           & \textbf{Train}   & \textbf{Valid}  & \textbf{Test}     \\ \midrule
		ISCXTor          &  614,575      & 450,001 & 82,287 & 82,287    \\
		ISCXVPN        &  492,598      & 443,337  & 24,631  &  24,630    \\
		CHNAPP    & 1,287,303    & 1,158,520 & 64,391 & 64,392   \\
		\bottomrule
	\end{tabular}
\end{table}

\begin{table*}[ht]
	\centering
	\caption{Comparison results on public and collected datasets. $^\dagger$ indicates reproduced results. Effi. metric means that the number of samples that the model predicts per second. Ma-F1 and Mi-F1 display using percentages (\%).}
	\label{tab:main}
	\resizebox{\linewidth}{!}{
		\begin{tabular}{c|ccc|ccc|ccc}
			\toprule
			\multicolumn{1}{c|}{\multirow{2}{*}{Methods}}   & \multicolumn{3}{c|}{ISCXVPN} & \multicolumn{3}{c|}{ISCXTor} & \multicolumn{3}{c}{CHNAPP}   \\ 
			\multicolumn{1}{c|}{} & Ma-F1  & Mi-F1 & Effi. & Ma-F1 & Mi-F1 & Effi. & Ma-F1 & Mi-F1 & Effi. \\ \midrule \midrule
			APPS  \cite{DBLP:journals/tifs/TaylorSCM18}    & 81.68   & 81.33   & -   & -   & 49.68   & -  & -   & - & - \\
			SMT  \cite{DBLP:conf/iwqos/ShenZZ0DL19}      & 76.31   & 76.00   & -   & -   & 31.54   & - & 34.77   & 45.62 & -  \\ \midrule
			Deep Packet \cite{DBLP:journals/soco/LotfollahiSZS20}  & 65.06   & 66.70   & -   & -   & 26.81   &  & 31.28   & 41.07 & -  \\
			TR-IDS \cite{DBLP:journals/scn/MinLLCC18}    & 54.98   & 57.66   & -   & -   & 20.75   & - & -   & - & -  \\
			HEDGE \cite{DBLP:journals/tifs/CasinoCP19}     & 29.67   & 34.00   & -   & -   & 15.08   & - & 25.67   & 36.35 & - \\
			3D-CNN \cite{DBLP:conf/infocom/ZhangLYW20}    & 54.34   & 57.00   & -   & -   & 33.96   & - & 38.74    & 51.12 & - \\ \midrule
			BERT \cite{DBLP:conf/naacl/DevlinCLT19}$^\dagger$    & 88.45  & 88.73   & 41.04   & 61.84        & 92.25   & 58.33  & 56.94    & 84.08  & 32.88 \\ 
			ET-BERT \cite{DBLP:conf/www/LinXGLSY22}$^\dagger$    & 94.25  & 96.05   & 10.83   & 95.12        & 97.86   & 18.99 & 94.08   & 96.09 & 20.14 \\ 
			YaTC  \cite{zhao2023yet}$^\dagger$  & 96.67  & 98.04   & 23.68   & 96.30  & 98.15   & 23.07 & 90.44   & 93.50 & 29.41 \\ 
			PacRep \cite{DBLP:conf/kdd/MengWMLLZ22} $^\dagger$    & 99.32   & 99.13   & 22.75   & 95.02   & 96.38   & 36.76 & \textbf{95.10}   & 98.75 & 44.46 \\ \midrule
			EETS (Ours)  & \textbf{99.89}   & \textbf{99.88}   & \textbf{124.01}   & \textbf{97.56}   & \textbf{98.87}  & \textbf{171.43}  & {94.95}   & \textbf{99.14}   & \textbf{82.11} \\ \bottomrule
	\end{tabular}}
\end{table*}

\section{Experiments}
\subsection{Public Dataset}
First of all, the proposed EETS and comparison methods are evaluated using two public traffic datasets \cite{DBLP:conf/icissp/Draper-GilLMG16,DBLP:conf/icissp/LashkariDMG17}. Specifically, the ISCXVPN \cite{DBLP:conf/icissp/Draper-GilLMG16} dataset predominantly consists of network traffic data using Virtual Private Networks (VPNs), with each packet labeled by application, such as Gmail, Facebook, and others. The ISCXTor \cite{DBLP:conf/icissp/LashkariDMG17} dataset utilizes The Onion Router (Tor) to enhance communication privacy and includes 13 categories of application classification, like Bittorent, Netflix, Youtube and so on. Detailed statistical information on the number of packets used for training, validation, and testing is provided in Table \ref{tab:stats}.

\subsection{Collected Dataset}
To further conduct an effective and efficient comparison between the SOTA models and EETS, we conduct a real-world dataset collection manually, named CHNAPP. This dataset comprises network traffic data from six widely-used Internet applications: Weibo, WeChat, QQMail, TaoBao, Youku, and QQMusic. By incorporating these popular applications, CHNAPP provides a robust basis for evaluating and contrasting the performance of EETS with leading models in real-world scenarios. This comprehensive collection ensures that the comparison is both realistic and relevant. The dataset will be released in the future, and the statistical dataset information can be seen in Table \ref{tab:stats}.

\subsection{Evaluation Metrics}
Evaluation metrics are designed from two perspective. For effectiveness, the category distribution in these tasks varies, with some being balanced and others imbalanced. To accurately assess performance, both Macro F1 (Ma-F1) and Micro F1 (Mi-F1) scores are utilized. The Macro F1 score calculates the F1 score independently for each class and then takes the average, treating each class equally regardless of its size. This approach is particularly effective for evaluating model performance in scenarios with imbalanced class distributions, as it ensures that smaller classes are given the same consideration as larger ones. Conversely, the Micro F1 score aggregates the contributions from all classes to compute a global F1 score. This method places more emphasis on the performance of larger classes, providing a measure of overall accuracy. By considering both Macro F1 and Micro F1 scores, we can provide a more comprehensive evaluation. For the efficiency calculation, {Effi. (Efficiency) metric means that the number of samples that the model predicts per second.} The above evaluation metrics are the higher the value, the better.

\subsection{Experimental Setting}
For the experimental settings, the model parameters are detailed in three aspects. Firstly, for the DPC selector, the training data comprises approximately 100,000 samples, while the testing data consists of 10,000 samples. The training process runs for 3 epochs with a learning rate of 1e-5 and a batch size of 64. Next, for the plaintext classification model, the one-hot layer dimension is set to 100, and the dimension of the 1D-convolution layer is set to (1, 10). This model is trained for 1 epoch with a learning rate of 1e-5 and a batch size of 64. Finally, for the encrypted text classification model, the BERT layer dimension is set to 768. All other parameters remain consistent with the Pacrep model \cite{DBLP:conf/kdd/MengWMLLZ22}. Epoch is set to 5 , with a learning rate of 5e-6 and a batch size of 64. The Adam \cite{Kingma2014AdamAM} optimizer with the initial learning rate of $1$e-$5$ is utilized for training, and the Pytorch framework \cite{Paszke2017AutomaticDI} is adapted. The experiments are implemented on the Tesla A100-80G GPU.

\subsection{Baselines}
The baselines can be divided into three categories for comparison: (1) feature-based methods (i.e., APPS \cite{DBLP:journals/tifs/TaylorSCM18}, SMT \cite{DBLP:conf/iwqos/ShenZZ0DL19}); (2) byte-based methods (i.e., Deep Packet \cite{DBLP:journals/soco/LotfollahiSZS20}, TR-IDS \cite{DBLP:journals/scn/MinLLCC18}, HEDGE \cite{DBLP:journals/tifs/CasinoCP19}, 3D-CNN \cite{DBLP:conf/infocom/ZhangLYW20}); and (3) PLMs-based methods (i.e., BERT \cite{DBLP:conf/naacl/DevlinCLT19}, PacRep \cite{DBLP:conf/kdd/MengWMLLZ22}, ET-BERT \cite{DBLP:conf/www/LinXGLSY22}, YaTC \cite{zhao2023yet}). In this paper, PLMs-based methods are main comparison points due to their competitive performance. More details of these compared baselines can be seen in Appendix \ref{app:baseline} because of the space limitation.

%

\begin{figure*}
	\centering
	\begin{tabular}{cc}
		\includegraphics[width=0.40\textwidth, height=0.35\textwidth]{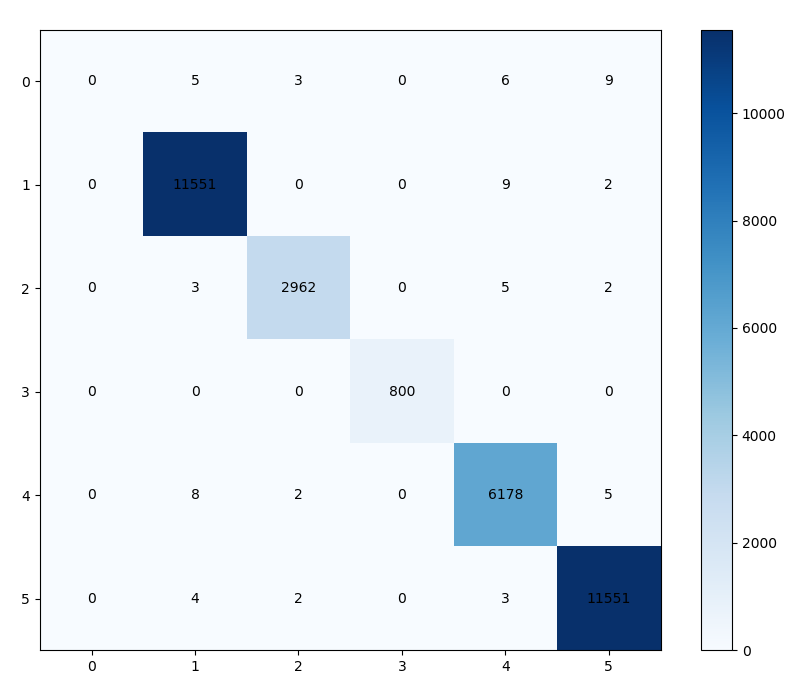} &
		\includegraphics[width=0.40\textwidth, height=0.35\textwidth]{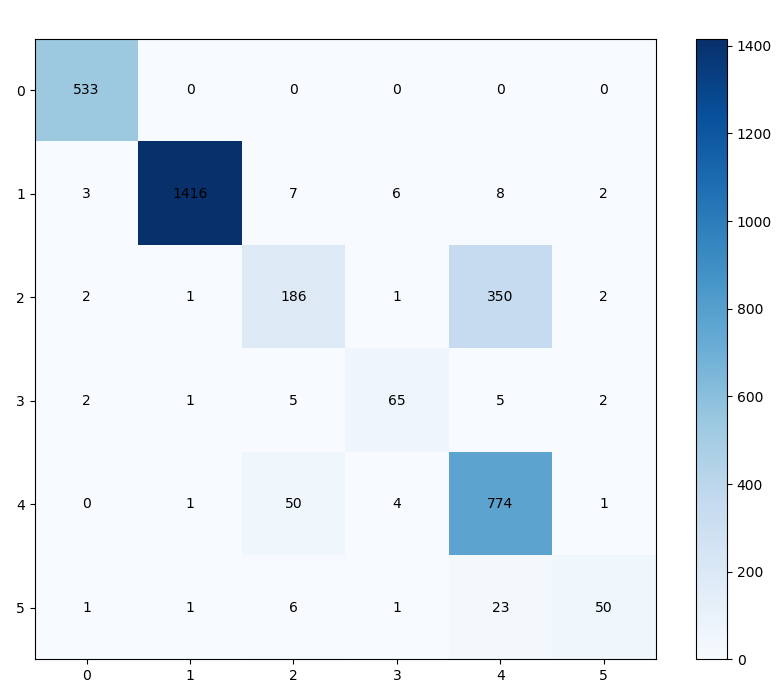} \\
		\textbf{(a) Confusion matrix on plaintext classifiable setting.} & \textbf{(b) Confusion matrix on plaintext non-classifiable setting.} \\
	\end{tabular}
	\caption{Confusion matrix visualization on two different settings on CHNAPP dataset. 0: QQMail, 1: QQMusic, 2: Youku, 3: TaoBao, 4: WeChat, 5: Weibo, where the proportion of QQMail data is relatively small.}
	\label{fig:led_vs_our}
\end{figure*}

\begin{table*}[ht]
	\centering
	\caption{Comparison results on different settings. $^\dagger$ indicates reproduced results. }
	\label{tab:diff}
	\resizebox{\linewidth}{!}{
		\begin{tabular}{c|ccc|ccc|ccc}
			\toprule
			\multicolumn{1}{c|}{\multirow{2}{*}{Methods}}   & \multicolumn{3}{c|}{ISCXVPN} & \multicolumn{3}{c|}{ISCXTor} & \multicolumn{3}{c}{CHNAPP}   \\ 
			\multicolumn{1}{c|}{} & Ma-F1  & Mi-F1 & Effi. & Ma-F1 & Mi-F1 & Effi. & Ma-F1 & Mi-F1 & Effi. \\ \midrule \midrule
			PacRep \cite{DBLP:conf/kdd/MengWMLLZ22} $^\dagger$    & 99.32   & 99.13   & 22.75   & 95.02   & 96.38   & 36.76 & {95.10}   & 98.75 & 44.46 \\ \midrule
			Plaintext Classifiable  & {99.98}   & {99.97}   & {304.41}   & {97.16}   & {99.97}   & {314.40}  & {79.85}   & {99.89}   & {103.40} \\
			Plaintext non-Classifiable  & {77.62}   & {94.22}   & {19.78}   & {86.48}   & {92.75}  & {48.03}  & {79.58}   & {86.18}   & {18.27} \\
			\textbf{Overall}  & {99.89}   & {99.88}   & {124.01}   & {97.56}   & {98.87}  & {171.43}  & {94.95}   & {99.14}   & {82.11} \\ \bottomrule
	\end{tabular}}
\end{table*}

\subsection{Main Results}
Considering the different techniques employed by feature-based methods, byte-based models, and PLMs-based methods, we focus our reproduction only on PLMs-based methods to ensure a fair comparison of time overheads and provide time efficiency metric (Effi.) for these methods. In addition, to highlight the effectiveness of PLMs-based methods, we compare their Macro-F1 and Micro-F1 scores with those of feature-based methods and byte-based models. Due to the relatively poor accuracy of feature-based methods and byte-based models, we do not compare their time overheads with our approach. Instead, we focus on comparing the time efficiency of our method with the current SOTA models, i.e. PLMs-based methods, to ensure both effectiveness and efficiency.

As shown in Table \ref{tab:main}, the conclusions drawn from this comparison are as follows: (1) In terms of effectiveness , feature-based methods and byte-based models perform comparably, while PLMs-based methods demonstrate further improvement. Notably, our proposed EETS model achieves SOTA results across almost all evaluation metrics, illustrating the superiority of the two-stage framework, which adaptively makes a balance between the plaintext and encrypted text. It is noteworthy that although EETS may obtain slightly lower performance in certain metrics, such as a 0.15\% reduction in the Macro-F1 score on the CHNAPP dataset, the model achieves approximately a 2 to 4 times improvement in efficiency. This significant enhancement in time efficiency validates that the proposed method not only maintains robust effectiveness but also substantially improves the prediction time efficiency of the model. 

(2) From a efficiency perspective, EETS exhibits a notable improvement across three datasets. Specifically, when compared to the PacRep model, a BERT-based model fine-tuned directly under supervised learning, the EETS model shows clear advantages (about 2 to 5 times improvement). On the other hand, ET-BERT and YaTC incorporate an additional pre-training phase before fine-tuning, resulting in greater overhead compared to directly fine-tuning BERT. 

Overall, the analysis of both effectiveness and efficiency demonstrates the superiority of the EETS framework and underscores the necessity of analyzing both plaintext and encrypted text.

\subsection{Results on Different Settings}
To perform a fine-grained analysis of our model's final performance (i.e., to understand why the model performs well), we conduct a detailed comparison in plaintext classifiable and non-classifiable settings. The backbone of our model is Pacrep. As shown in Table \ref{tab:diff}, our model achieves overall optimal performance across three datasets, excelling in both effectiveness and efficiency metrics.

Specifically, in the plaintext classifiable setting, our model EETS demonstrates superior prediction efficiency, outperforming baseline models by approximately 3 to 15 times while maintaining decent accuracy. This significant improvement underscores the model's capability to handle classifiable plaintext data efficiently. In addition, in the plaintext non-classifiable setting, the model's accuracy is relatively lower. This phenomenon is likely due to the inherent difficulty of classifying these difficult samples. Despite this, our model still achieves almost SOTA results in terms of overall performance. For further analysis of the Macro-F1 scores, refer to \textbf{Section \ref{confuse} Confusion Matrix Visualization}.

\begin{table*}[ht]
	\centering
	\caption{Results of DPC selector analysis on different datasets. Num.: the number of the samples.}
	\label{tab:dpc}
	\resizebox{\linewidth}{!}{
		\begin{tabular}{c|ccc|ccc|ccc}
			\toprule
			\multicolumn{1}{c|}{\multirow{2}{*}{Methods}}   & \multicolumn{3}{c|}{ISCXTor} & \multicolumn{3}{c|}{ISCXVPN} & \multicolumn{3}{c}{CHNAPP}   \\ 
			\multicolumn{1}{c|}{} & Ma-F1  & Mi-F1 & Num. & Ma-F1 & Mi-F1 &  Num. & Ma-F1 & Mi-F1 &  Num. \\ \midrule \midrule
			DPC Selector (ours)    & 97.56   & 97.56   & 82,287   & 89.65   & 99.61   & 24,630 & {92.45}   & 98.60 & 64,392 \\  \bottomrule
	\end{tabular}}
\end{table*}

\begin{figure*}
	\centering
	\begin{tabular}{cc}
		\includegraphics[width=0.45\textwidth, height=0.26\textwidth]{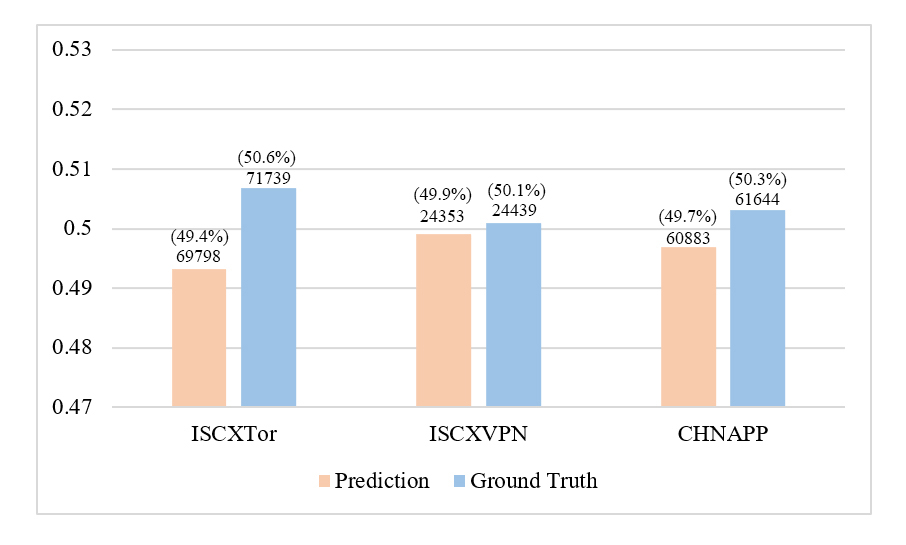} &
		\includegraphics[width=0.45\textwidth, height=0.26\textwidth]{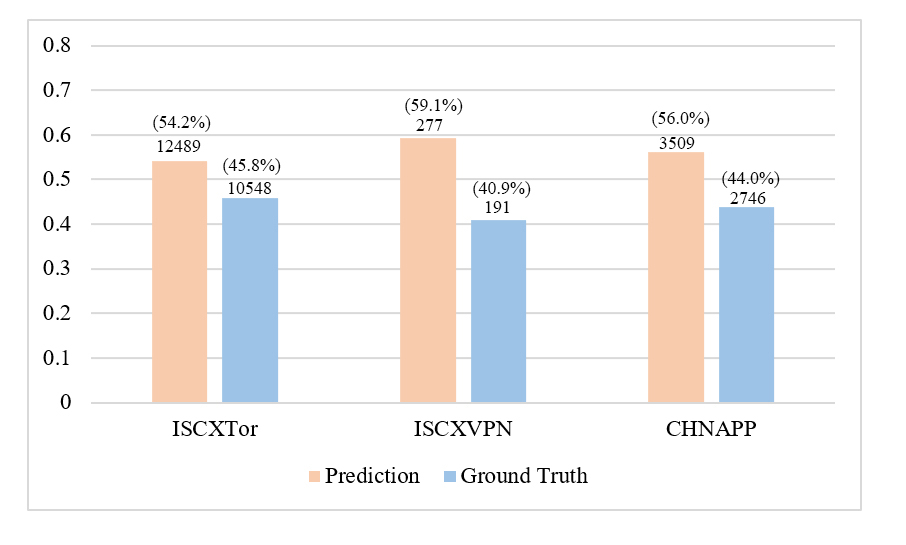} \\
		\textbf{(a) Testing sample size analysis on setting one.} & \textbf{(b) Testing sample size analysis on setting two.} \\
	\end{tabular}
	\caption{Testing  sample size analysis on two different settings.}
	\label{fig:sample}
\end{figure*}

\section{Analysis}
\subsection{Confusion Matrix Visualization}
\label{confuse}
In Table \ref{tab:diff}, we have made a new discovery. Horizontally examining the three datasets, particularly CHNAPP, we observe that the Macro-F1 scores under both settings are relatively low. However, the final Macro-F1 score reaches an impressive 94.95\%. To analyze this phenomenon, we conduct a detailed statistical analysis of the results for different classes within CHNAPP and visualize the confusion matrix in Figure \ref{fig:led_vs_our}. Figures \ref{fig:led_vs_our} (a) and (b) represent the plaintext classifiable and non-classifiable settings, respectively.
From Figure \ref{fig:led_vs_our} (a), it is evident that most samples are accurately classified. Specifically, the accuracy for the last four classes is nearly 99\%, although there is some misclassification in the \textit{QQmail} class, which leads to a lower Macro-F1 score (79.85\%), while the Micro-F1 score reaches an outstanding 99.89\%. Additionally, in Figure \ref{fig:led_vs_our} (b), we observe: (1) The classification difficulty in setting two is higher compared to the plaintext classifiable setting. Specificially, there exists lower performance on the \textit{Youku} class samples, while other categories still achieve good levels of accuracy. (2) The \textit{QQmail} class samples are all correctly classified, further proving that the introduction of encrypted text effectively helps the model in classification. This makes up for the misclassification issue of the \textit{QQmail} class samples observed in setting one.

All in all, the proposed model's performance metrics on the CHNAPP dataset are decent, with a Macro-F1 of 94.95\% and a Micro-F1 of 99.14\%. This observation holds true for other datasets as well, indicating consistent performance improvements across different scenarios.

\subsection{DPC Selector Analysis}
\label{sec:dpc}
The proposed method is a two-stage approach where the outcome of the first stage serves as the input for the second stage. Consequently, the performance of the model in the first stage significantly impacts the results of the second stage. To further analyze the model's performance and demonstrate the effectiveness of the two-stage approach, this section presents the DPC selector analysis. Specifically, we examine the performance of the DPC selector in the first stage. As shown in the Table \ref{tab:dpc}, the Micro-F1 scores exceed 97\% across all three datasets, indicating that almost all data can be correctly selected, thereby ensuring robust performance in the second stage. 

Regarding the Macro-F1 metric, the score for the ISCXTor dataset is slightly lower at 89.65\%, likely due to data imbalance. However, the Micro-F1 score for ISCXTor reaches 99.61\%, making the impact on the final classification results negligible. Additionally, the Macro-F1 scores remain high for the other two datasets. All in all, this experimental analysis confirms the effectiveness of the proposed DPC selector, therefore guaranteeing the accurate traffic classification in the second stage.

\subsection{Testing Sample Size Analysis}
To further validate the performance of the DPC selector and the reliability of plaintext classification, we conduct a sample size analysis under both plaintext classification and non-classification settings. As shown in Figure \ref{fig:sample} (a), in setting one, the ratio of samples predicted to be plaintext-classifiable closely matches the actual samples, with both ratios covering around 50\%. Similarly, in setting two, the ratio is maintained between 45\% and 55\%, with a slightly larger variance observed in the ISCXVPN dataset. These results further demonstrate the effectiveness of the DPC selector.

In a cross-comparison, as illustrated in Figures \ref{fig:sample} (a) and (b), the number of samples predicted to be plaintext-classifiable accounts for a large proportion. Specifically, in the ISCXVPN dataset, the results reach nearly 100 times (24,439/191), while in the other two datasets, the results range from 7 to 30 times. This strongly indicates that relying on plaintext information can address the majority of traffic classification scenarios. And encrypted data can also aid in improving classification accuracy alongside plaintext. Moreover, the simplicity and readability of plaintext further enhance the model's prediction efficiency, providing that the motivation behind this paper is both effective and necessary.

\section{Conclusion}
In this paper, we propose a simple but Efficient and Effective Two-Stage approach (EETS) to balance plaintext and encrypted text for traffic classification. Moreover, we make the first attempt to analyze the impact of plaintext and encrypted text on model performance and time efficiency,
which provides a new insight to make an analysis on the plaintext and encrypted text in network traffic field. For experiments, EETS is very effective and efficient, which outperforms the previous methods on two public datasets and one real-world dataset collected by ourselves. The analyses also demonstrate the necessity and interpretability. 
For future work, the design of modules in the EETS could be considered for further research.


\appendix
\section{Appendix}
\subsection{Details of Baselines}
\label{app:baseline}
The details of utilized baselines can be seen in the following.
\begin{itemize}
	\item APPS \cite{DBLP:journals/tifs/TaylorSCM18}, which selects statistical features, and uses a random tree classifier for application classification.
	\item SMT \cite{DBLP:conf/iwqos/ShenZZ0DL19}, which fuses features of different dimensions by a kernel function for decentralized applications fingerprinting.
	\item Deep Packet \cite{DBLP:journals/soco/LotfollahiSZS20}, which integrates both feature extraction and classification phases to train the stacked autoencoder and CNN for application classification.
	\item 3D-CNN \cite{DBLP:conf/infocom/ZhangLYW20}, which leverages the byte data as input for 3D-CNN and can classify packets from both known and unknown patterns.
	\item BERT \cite{DBLP:conf/naacl/DevlinCLT19}, which is based on Transformer architecture and has achieved noticeable performance on various NLP tasks.
	\item PacRep \cite{DBLP:conf/kdd/MengWMLLZ22}, which utilizes the BERT as an encoder and uses contrastive loss to optimize the learned packet representations with multi-task learning.
	\item ET-BERT \cite{DBLP:conf/www/LinXGLSY22}, which pre-trains deep contextualized datagram-level representation from large-scale raw traffic bytes.
	\item YaTC \cite{zhao2023yet}, which introduces a masked autoencoder based traffic transformer with multi-level flow representation and then performs pre-training process for traffic classification.
\end{itemize}

\bibliographystyle{ACM-Reference-Format}
\balance
\bibliography{sample-base}

\end{document}